\documentstyle[preprint]{ptptex}

\newcommand{\bl}[1]{\mbox{\boldmath{$ #1$}}}
\newcommand{\epsl}{\varepsilon}

\title{
Statistics of the Composite System
}

\author{
Hitoshi {\sc Ito}\footnote{
 E-mail address: itoh@phys.kindai.ac.jp}
}

\inst{
Department of Physics, Faculty of Science and
 Engineering, Kinki University, Higashi-Osaka, 577 }

\recdate{
}

\abst{
The space-like asymptotic limit of the bilocal composite field of the state consisting of a nucleus and an electron is studied. It is shown that the resulting local field of an atom satisfies the proper commutation relations in the sub-Fock-space of the atom.
}

\begin{document}

\maketitle

More than forty years ago, composite systems in the quantum field theory was investigated\cite{Zimmermann58}. 
One starts from a local scalar field, say $A(x)$, and assumes the existence of discrete eigenvalues $m^2$ and $M^2$ of the 4-momentum squared $P^2$, where $m$ is the mass of the original field $A$.
If $\langle A(x)A(y)\mid P^2=M^2\rangle \neq 0$ there may be a composite(bound)  state of the mass $M$. Then, one defines a bilocal field
\begin{equation}
  B(x,\epsl)=TA(x+\epsl)A(x-\epsl)        \label{a1}
\end{equation}
representing this state, where $T$ denotes the time-ordered product.
Zimmermann showed, under some mathematical assumption, that the asymptotic($t\to\pm\infty$) field of $B$ satisfies the proper commutation relation in the limit $\epsl\to 0$, if it is suitablly normalized.
 One can infer, from this construction, that composite systems can be described  by the local field operators and there are no differences between elementary and composite particles in constructing the $S$ matrix elements.
An idea of the boot strapping(nuclear democracy) emerged from this observation.

However, the successes of the gauge theories have changed drastically the framework of the elementary particle theory. 
The quantum theory of field has revived and one gradually recognized the hierarchy structure of the gauge interactions. 
A field theory now becomes an effective theory for each class of the hierarchy.

On the other hand, there is another hierarchy of compositeness in nature, which was revealed through success of the composite models of elementary particles. The hierarchy here consists of the classes of quark, hadron, neucleus, atom and so on, although we are not sure if the class of quark is the deepest.
 The level of a class is specified by the energy scale indicating the limit of applicability of the theory governing it\footnote{The energy scale can be replaced by the length scale in some case.}.
We, further, believe that the theory is a quantum field theory and there exist elementary fields for each class, which are constructed from the elementary fields of the deeper class.
The most instructive example of this interpolating mechanism may be provided by considering the class of atom. 
An atom is a composite system which consists of a nucleus field and an electron field interacting through the photon field. Then, we construct the atom field as a composite field of the elementary fields of the deeper class. In this respect, we are specially interested in the statistic property of the composite system, since we are hardly convinced of it by the particle quantum mechanics.
 For definiteness and simplicity, we restrict ourselves to the hydrogen atom in the following. There is also another reason to take up it: The nucleus of it is a Fermi particle and it changes to the Bose particle by acquiring another Fermion. This is mysterious from the view point of the particle quantum mechanics.

Let us first introduce the composite field $\Psi$ through the equation
\begin{equation}
   \Psi(x_1,x_2)=T\psi(x_1)\phi(x_2)       \label{a2}
\end{equation}
where $\psi(x_1)$  and $\phi(x_2)$ are the elementary fields of the nucleus and the electron, respectively, in the Heisenberg picture. We, then, define the Bethe Salpeter amplitude $\langle 0|\Psi(x_1,x_2)|2\rangle$ for the two-particle states and obtain the state of the H-atom by solving the BS equation for it. The nonrelativistic approximation suffices for the present purpose. And we are interested in only the wave function of the ground state, the Fourier transform of which is denoted by $g_{mm'}(\bl{k}_1,\bl{k}_2)$, where $m$ and $m'$ are the indices of the spin of the nucleus and the electron respectively. We note, however, that these indices are dummy since the spin of the nucleus is frozen in the nonrelativistic approximation and therefore the spin freedom of the electron can be neglected in the ground $S$ state. 

We next reconstruct the composite field operator by including the bound-state amplitude. If the total and the relative momenta are $K=(K_0,\bl{K})$ and $k=(k_0,\bl{k})$ respectively, the contribution of the bound state to the annihilation part is given by
\begin{eqnarray}
 \lefteqn{ \Psi(X,x) = Ce^{iKX}\sum_{mm'}\int d^3k} \label{a3} \\
   & & \times g_{mm'}(\bl{K},\bl{k})
    a_m(\eta_1\bl{K}-\bl{k})b_{m'}(\eta_2\bl{K}+\bl{k})
                            \exp{(i\bl{k}\cdot\bl{x})},
                     \quad \eta_1+\eta_2=1,      \nonumber
\end{eqnarray}
where $C$ is a normalization factor and $a_m(\bl{k}_1)$ and $b_{m'}(\bl{k}_2)$ are the annihilation operators, respectively, for the nucleus and the electron which satisfy the anti-commutation relations
\begin{equation}
    \{a_m(\bl{k}_1),a^\dagger_{m'}(\bl{k}_2)\}=
      \delta_{mm'}\delta(\bl{k}_1-\bl{k}_2), \quad \mbox{etc.}.   \label{a4}
\end{equation}
Now, when we observe the stable atom from some great distance we can neglect the scale of the relative coordinate $\bl{x}$ and the atom is represented by the wave function at the origin\footnote{
The wave function at the origin becomes a divergent quantity in some class of the relativistic equation. We have to renormalize it in such a case\cite{Ito82}.
}.
 We call the neglect of the relative coordinate the space-like asymptotic limit.

The bound state is represented by a local field operator in the space-like asymptotic limit, the annihilation part of which is given by
\begin{equation}
  \Psi(X,0)=CA(\bl{K})e^{iKX},  \quad 
   A(\bl{K})=\int g(\bl{K},\bl{k})a(\eta_1\bl{K}-\bl{k})b(\eta_2\bl{K}+\bl{k})d^3k,  
                               \label{a5}
\end{equation}
where the dummy spin indices are omitted. $A(\bl{K})$ satisfies commutation relations
\begin{equation}
   [A(\bl{K}),A(\bl{K}')]=[A^\dagger(\bl{K}),A^\dagger(\bl{K}')]=0
\end{equation}
and
\begin{eqnarray}
 \lefteqn{ [A(\bl{K}),A^\dagger(\bl{K}')] = \delta(\bl{K}-\bl{K}')} 
      \label{a61} \\ & & 
   -\int d^3k g(\bl{K},\bl{k})g^*(\bl{K}',\eta_2\bl{K}-\eta_2\bl{K}'+\bl{k})
 a^\dagger(\bl{K}'-\eta_2\bl{K}-\bl{k})a(\eta_1\bl{K}-\bl{k}) \nonumber \\ & &
 -\int d^3k g(\bl{K},\bl{k})g^*(\bl{K}',\eta_1\bl{K}'-\eta_1\bl{K}+\bl{k})
 b^\dagger(\bl{K}'-\eta_1\bl{K}+\bl{k})b(\eta_2\bl{K}+\bl{k}) \nonumber
\end{eqnarray}
where the normalization of the mometum-space wave function is assumed to be 1. The spectral condition forbids the last two terms in the right hand side to have matrix elements within the subspace of the bound state. We therefore neglect them and reach the proper commutation relations for the asymptotic atom field. 

It is not persuadable to interpret the statistic property of the bound state by using the particle quantum mechanics. Instead, we should first conceive of the composite field consisting of the constituent fields, as shown in preceding consideration. Translation of it into the language of the particle quantum mechanics may be that `the constituents loose their individuality in the bound state and behave as a quantum mechanical unity'\cite{Ito98}.

Finally, we mention the Cooper pair in superconductors which may be a composite system in the top class of the hierarchy. Although it is not confined in a space region, the similar arguments leading to (\ref{a61}) can be made if we normalize the pair wave function in an appropriate space volume.


\section*{Acknowledgements}

The author has been stimulated by discussions on BEC in the workshop "Thermo Field Dynamics" held at RIFP in the mid summers of 1999 and 2000. He would like to thank the organizers and participants with whom he enjoyed having discussions.


\end{document}